\documentclass{article}

\usepackage{cite}
\usepackage{graphicx}

\begin{document}

\date{}
\title{A quantum-mechanical anharmonic oscillator with a most interesting spectrum}
\author{Paolo Amore \thanks{%
paolo.amore@gmail.com} \\
%EndAName
Facultad de Ciencias, Universidad de Colima, \\
Bernal D\'{\i}az del Castillo 340, Colima, Colima, Mexico\\
Francisco M. Fern\'andez \thanks{%
fernande@quimica.unlp.edu.ar}\\
INIFTA (CONICET, UNLP), Divisi\'{o}n Qu\'{i}mica Te\'{o}rica,\\
Blvd. 113 y 64 (S/N), Sucursal 4, Casilla de Correo 16,\\
1900 La Plata, Argentina}
\maketitle

\begin{abstract}
We revisit the problem posed by an anharmonic oscillator with a potential
given by a polynomial function of the coordinate of degree six that depends
on a parameter $\lambda $. The ground state can be obtained exactly and its
energy $E_{0}=1$ is independent of $\lambda $. This solution is valid only
for $\lambda >0$ because the eigenfunction is not square integrable
otherwise. Here we show that the perturbation series for the expectation
values are Pad\'{e} and Borel-Pad\'{e} summable for $\lambda >0$. When $%
\lambda <0$ the spectrum exhibits an infinite number of avoided crossings at
each of which the eigenfunctions undergo dramatic changes in their spatial
distribution that we analyze by means of the expectation values $\langle
x^{2}\rangle $.
\end{abstract}

\section{Introduction}

\label{sec:intro}

Some time ago Herbst and Simon\cite{HS78} discussed some interesting and
baffling features of two one-dimensional Hamiltonians. In one of them, $%
H^{(2)}(g)=p^{2}+x^{2}-1+g^{4}x^{6}+2g^{2}x^{4}-3g^{2}x^{2}$, the exact
ground-state energy is $E^{(2)}(g)=0$ and the coefficients of the
perturbation series $\sum a_{n}^{(2)}g^{2n}$ vanish for all $n>0$. However,
the perturbation series for the eigenvector $\Omega ^{(2)}(g)$ is divergent
at least in the norm sense. The related oscillator $H^{(3)}(g)=H^{(2)}(ig)$
is most interesting because $0<E^{(3)}(g)<D^{\prime }\exp (-c/g^{2})$. Its
potential has three wells and there is a kind of asymptotic degeneracy of
expected states.

Those models are particular cases of the so-called quasi-exactly solvable
Schr\"{o}dinger equations\cite{T16} (and references therein). In fact,
Turbiner\cite{T16} chose the closely related potential $%
V_{0}(x;a,b)=a^{2}x^{6}+2abx^{4}+\left( b^{2}-3a\right) x^{2}-b$ for the
discussion of the most interesting problem of phase transition.

The purpose of this paper is the analysis of the spectra of $H^{(2)}$ and $%
H^{(3)}$ because they exhibit several interesting features that may not
emerge so clearly from the remarkable theoretical analysis carried out by
Herbst and Simon\cite{HS78} and Turbiner\cite{T16}. Present results are
shown in section~\ref{sec:model} and conclusions in section~\ref
{sec:conclusions}.

\section{The model}

\label{sec:model}

For simplicity, here we rewrite the Hamiltonian proposed by Herbst and Simon%
\cite{HS78} as
\begin{equation}
H(\lambda )=H^{(2)}\left( 2\sqrt{\lambda }\right) +1=p^{2}+x^{2}-12\lambda
x^{2}+8\lambda x^{4}+16\lambda ^{2}x^{6}.  \label{eq:H_ANHO}
\end{equation}
It exhibits an exact ground-state eigenfunction
\begin{equation}
\varphi (x)=\exp \left( -x^{2}/2-\lambda x^{4}\right) ,
\label{eq:psi_0_ANHO}
\end{equation}
with eigenvalue $E_{0}=1$. This solution is only valid for $\lambda \geq 0$
because it is not square integrable for negative values of $\lambda $.

\begin{figure}[tbp]
\begin{center}
\bigskip\bigskip\bigskip \includegraphics[width=9cm]{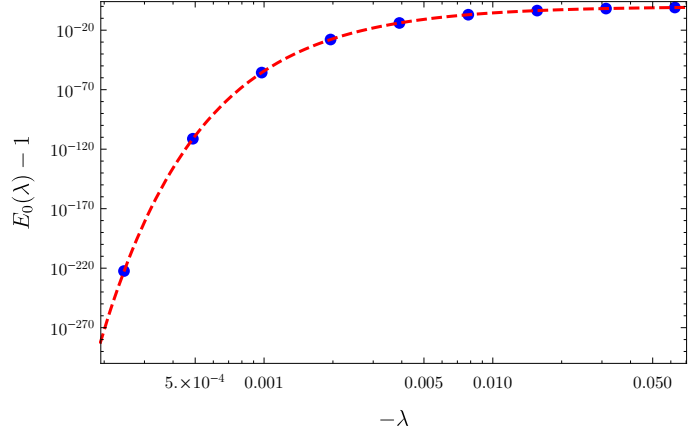}
\end{center}
\caption{$E_0(\lambda)-1$ calculated by means of the Rayleigh-Ritz
variational method (dashed red line) and its least-square fitting using
equation (\ref{eq:Delta_ANHO}) (blue points)}
\label{Fig:LN_DELTA}
\end{figure}

In principle, one expects the eigenfunctions and eigenvalues of $H(\lambda )$
to have perturbation expansions about $\lambda =0$ of the form
\begin{eqnarray}
\psi _{n}(x) &=&\sum_{p=0}^{\infty }\psi _{n}^{(p)}(x)\lambda ^{p},
\nonumber \\
E_{n} &=&\sum_{p=0}^{\infty }E_{n}^{(p)}\lambda ^{p}.
\label{eq:PT_series_general}
\end{eqnarray}
For the normalized ground-state eigenfunction we have
\begin{eqnarray}
\psi _{0}(x) &=&\frac{\varphi (x)}{\sqrt{\left\langle \varphi \right| \left.
\varphi \right\rangle }}  \nonumber \\
&=&\frac{1}{\sqrt{2}\pi ^{1/4}}e^{-x^{2}/2}\left[ 1+\frac{1}{4}\left(
3-4x^{4}\right) \lambda \right.   \nonumber \\
&&\left. +\frac{1}{32}\left( 16x^{8}-24x^{4}-183\right) \lambda ^{2}+\ldots
\right] ,  \label{eq:psi_0_series}
\end{eqnarray}
but all the perturbation corrections of the corresponding eigenvalue vanish (%
$E_{0}^{(j)}=0$, $j>0$) as mentioned above. Therefore, perturbation theory
fails to provide suitable values of $E_{0}(\lambda )$ when $\lambda <0$. The
reason is that this eigenvalue behaves asymptotically as\cite{HS78}
\begin{equation}
E_{0}(\lambda )-1\approx A|\lambda |^{B}e^{-C/|\lambda |},\;\lambda <0.
\label{eq:Delta_ANHO}
\end{equation}
Figure~\ref{Fig:LN_DELTA} shows that $E_{0}(\lambda )-1$ already behaves in
this way. A straightforward least-squares fitting for sufficiently small
values of $|\lambda |$ suggests that $A\approx 0.891$, $B=0$ (as argued by
Herbst and Simon\cite{HS78}) and $C=1/8$.

\begin{figure}[tbp]
\begin{center}
\bigskip\bigskip\bigskip \includegraphics[width=9cm]{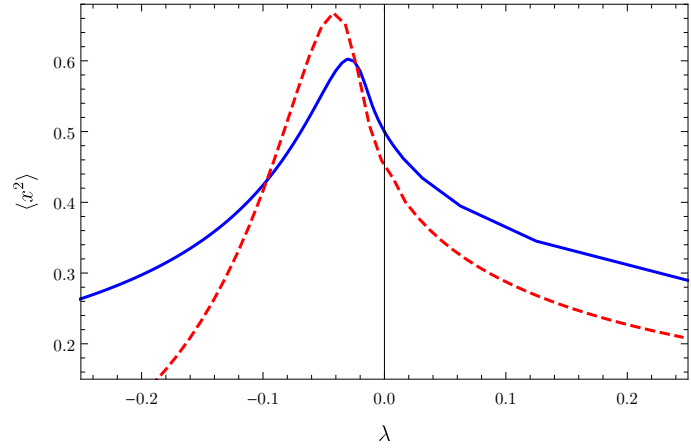}
\end{center}
\caption{Numerical $\langle x^2\rangle$ (blue, continuous line) and $\Re
S_B(\lambda)$ (dashed, red line) for the ground state of the oscillator (\ref
{eq:H_ANHO}) }
\label{Fig:X2MLONG}
\end{figure}

Although the perturbation series for the lowest eigenvalue converges for all
$\lambda $ that for its eigenfunction is divergent\cite{HS78}. As an
illustrative example consider the expectation value
\begin{equation}
\left\langle x^{2}\right\rangle =\frac{1}{2}-3\,\lambda +48\,{\lambda }%
^{2}-1188\,{\lambda }^{3}+39168\,{\lambda }^{4}-1604448\,{\lambda }%
^{5}+\ldots .
\end{equation}
for the ground state. In what follows we resort to the notation $%
X_{n}=\left\langle x^{2n}\right\rangle $ and $X_{n}^{(j)}$ for the
perturbation correction of order $j$. We can easily calculate the
perturbation corrections of $E_{n}^{(j)}$ and $X_{n}^{(j)}$ analytically to
any desired order by means of the hypervirial perturbation method\cite{F01}.
A least-squares fitting of the first $1000$ perturbation coefficients
enables us to estimate the asymptotic expansion
\begin{equation}
X_{1}^{(j)}=(-1)^{j}8^{j}j!\left[ f_{0}+\frac{f_{1}}{j+1}+\frac{f_{2}}{%
\left( j+1\right) ^{2}}+\ldots \right] ,\;j\gg 1,
\end{equation}
where
\begin{eqnarray*}
f_{0} &=&0.450158158079,\;f_{1}=-0.168809309279,\;f_{2}=-0.305966873069, \\
f_{3} &=&-0.869104178243,\;f_{4}=-3.78728795807,\;f_{5}=-22.6102214156.
\end{eqnarray*}
On keeping just the leading term $X_{1}^{(j)}\sim f_{0}(-1)^{j}8^{j}j!$ the
Borel sum yields
\begin{eqnarray}
S(\lambda ) &=&f_{0}\sum_{j=0}^{\infty }(-1)^{j}(8\lambda
)^{j}j!=f_{0}\int_{0}^{\infty }e^{-t}\sum_{j=0}^{\infty }(-1)^{j}(8\lambda
t)^{j}  \nonumber \\
S_{B}(\lambda ) &=&f_{0}\int_{0}^{\infty }\frac{e^{-t}}{1+8\lambda t}dt=f_{0}%
\frac{e^{\frac{1}{8\lambda }}}{8\lambda }\left[ \mathrm{Shi}\left( \frac{1}{%
8\lambda }\right) -\mathrm{Chi}\left( \frac{1}{8\lambda }\right) \right] ,
\end{eqnarray}
where
\begin{equation}
\mathrm{Shi}(x)=\int_{0}^{x}\frac{\sinh t}{t}dt,\;\;\mathrm{Chi}%
(x)=\int_{0}^{x}\frac{\cosh t}{t}dt.
\end{equation}
The Borel sum $S_{B}(\lambda )$ is complex for $\lambda <0$ and
\begin{equation}
\Im S_{B}(\lambda )\sim 0.176715|\lambda |^{-1}e^{-\frac{1}{8\lambda }%
},\;\;\lambda \rightarrow 0^{-}.
\end{equation}
Figure~\ref{Fig:X2MLONG} shows that the real part of $S_{B}(\lambda )$
exhibits a maximum for $\lambda <0$ like the actual value of $\left\langle
x^{2}\right\rangle $.

\begin{figure}[tbp]
\begin{center}
\bigskip\bigskip\bigskip \includegraphics[width=9cm]{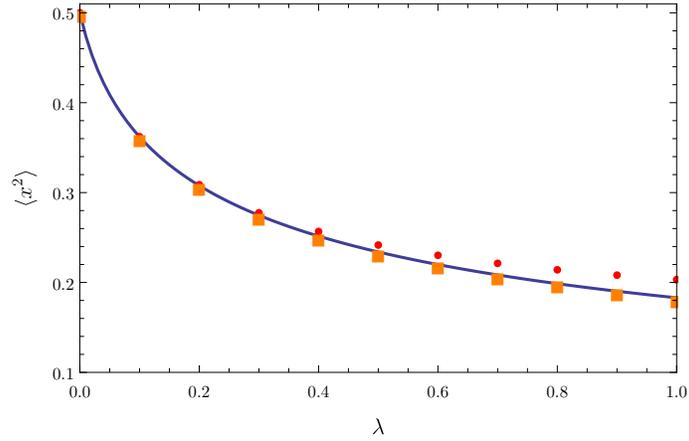}
\end{center}
\caption{Exact $\langle x^2 \rangle$ (solid line) for the ground state of
the oscillator (\ref{eq:H_ANHO}) and the $[6/6]$ Pad\'e (circles) and
Borel-Pad\'e (squares) sums of the perturbation series}
\label{Fig:X2M}
\end{figure}

The perturbation series originated in the expansion of a potential about one
of its minima can be shown to be non-Borel summable when the potential has
degenerate minima\cite{ZJ03}. It has been argued that in such a case the
imaginary part of the Borel sum is cancelled by the imaginary part of a
logarithmic term\cite{ZJ03}. In the present case the perturbation series are
Pad\'{e} and Borel-Pad\'{e} summable for $\lambda >0$ as shown in Figure~\ref
{Fig:X2M} for $\left\langle x^{2}\right\rangle $ (ground state). This figure
shows that the Borel summation improves the accuracy of the Pad\'{e}
approximant $[6/6](\lambda )$. However, both summation methods fail for $%
\lambda <0$.

The perturbation series for the excited states
\begin{eqnarray}
E_{1}(\lambda ) &=&3+12\,\lambda -144\,{\lambda }^{2}+4176\,{\lambda }%
^{3}-172800\,{\lambda }^{4}+8892288\,{\lambda }^{5}+\ldots  \nonumber \\
E_{2}(\lambda ) &=&5+48\,\lambda -864\,{\lambda }^{2}+36864\,{\lambda }%
^{3}-2194560\,{\lambda }^{4}+158810112\,{\lambda }^{5}+\ldots ,
\label{eq:E_j_series_ANHO}
\end{eqnarray}
are divergent; for example
\begin{equation}
E_{1}^{(j)}\sim (-1)^{j+1}\sqrt{j}8^{j}j!,
\end{equation}
was also obtained by numerical least-squares fitting of the analytical
perturbation corrections calculated by means of the hypervirial perturbation
method\cite{F01}.

\begin{figure}[tbp]
\begin{center}
\bigskip\bigskip\bigskip \includegraphics[width=9cm]{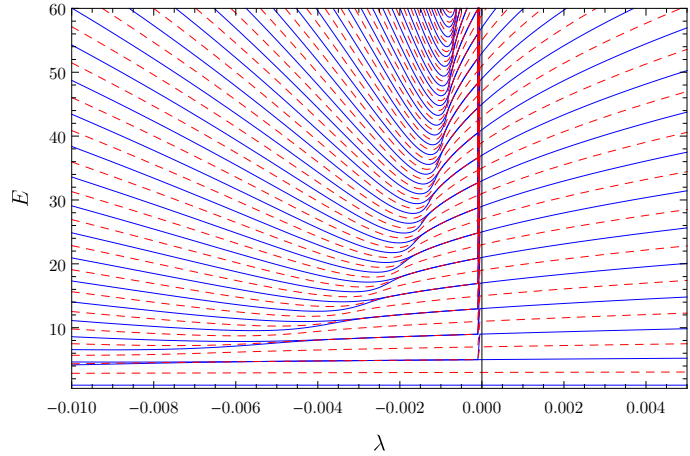}
\end{center}
\caption{Part of the spectrum of the anharmonic oscillator (\ref{eq:H_ANHO}%
). Even and odd states are denoted by continuous (blue) and dashed (red)
lines, respectively.}
\label{Fig:SPECTDET}
\end{figure}

Fig \ref{Fig:SPECTDET} shows the energy spectrum for small negative values
of $\lambda $. In order to understand its structure we should pay attention
to the form of the potential-energy function. When $0<\lambda <1/12$ the
potential is a single well and becomes a double well when $\lambda >1/12$,
but these cases are not relevant for present discussion. We just mention
them for completeness. When $\lambda <-1/36$ the potential is a single well;
when $-1/36<\lambda <0$ it exhibits three wells, one of them $V(0)=0$ at the
origin and the other two at $\pm x_{m}$, where
\begin{equation}
x_{m}^{2}=-\frac{\sqrt{36\lambda +1}+2}{12\lambda }=-\frac{1}{4\lambda }-%
\frac{3}{2}+\frac{27\lambda }{2}-243\lambda ^{2}+\ldots .
\label{eq:x_m_series}
\end{equation}
These side wells are separated from the central one by two barriers located
at $\pm x_{M}$ where
\begin{equation}
x_{M}^{2}=\frac{\sqrt{36\lambda +1}-2}{12\lambda }=-\frac{1}{12\lambda }+%
\frac{3}{2}-\frac{27\lambda }{2}+243\lambda ^{2}+\ldots .
\label{eq:x_M_series}
\end{equation}
Clearly the side wells move away from the origin as $\lambda \rightarrow
0^{-}$. The values of the potential at these stationary points are $V(0)=0$,
\begin{eqnarray}
V(x_{m}) &=&\frac{\left( \sqrt{36\lambda +1}+2\right) \left( \sqrt{36\lambda
+1}+36\lambda -1\right) }{54\lambda }=3+9\lambda -54\lambda ^{2}+\ldots ,
\nonumber \\
V(x_{M}) &=&\frac{\left( \sqrt{36\lambda +1}-2\right) \left( \sqrt{36\lambda
+1}-36\lambda +1\right) }{54\lambda }=-\frac{1}{27\lambda }+1-9\lambda
+\ldots .  \label{eq:V(x_m)_V(x_M)}
\end{eqnarray}
Note that the minima are bounded from below while the maxima increase
unboundedly. In the limit $\lambda \rightarrow 0^{-}$ we are left with a
harmonic oscillator. The curvatures of the minima and maxima tend to
constant values as $\lambda \rightarrow 0^{-}$
\begin{eqnarray}
V^{\prime \prime }(0) &=&2\left( 1-12\lambda \right)  \nonumber \\
V^{\prime \prime }(x_{m}) &=&\frac{8\left( 36\lambda +1+2\sqrt{36\lambda +1}%
\right) }{3}=8+192\lambda -864\lambda ^{2}+15552\lambda ^{3}+\ldots
\nonumber \\
V^{\prime \prime }(x_{M}) &=&\frac{8\left( 36\lambda +1-2\sqrt{36\lambda +1}%
\right) }{3}=-\frac{8}{3}+864\lambda ^{2}-15552\lambda ^{3}+\ldots
\label{eq:V"(0)_V''(x_m)_V''(x_M)}
\end{eqnarray}

\begin{figure}[tbp]
\begin{center}
\bigskip\bigskip\bigskip \includegraphics[width=9cm]{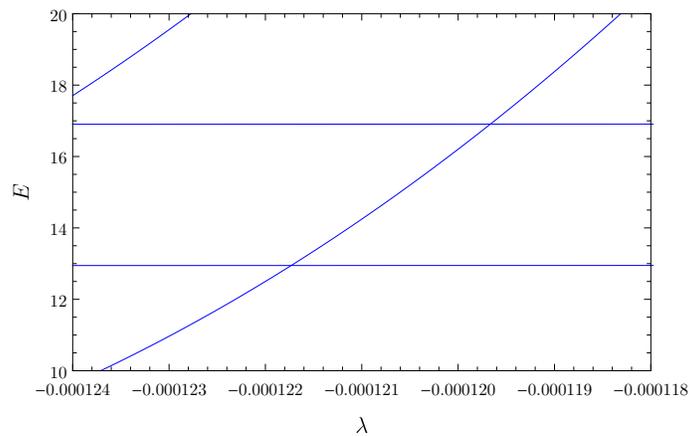}
\end{center}
\caption{Part of the spectrum of even states of the anharmonic oscillator (%
\ref{eq:H_ANHO})}
\label{Fig:DETEV}
\end{figure}

\begin{figure}[tbp]
\begin{center}
\bigskip\bigskip\bigskip \includegraphics[width=9cm]{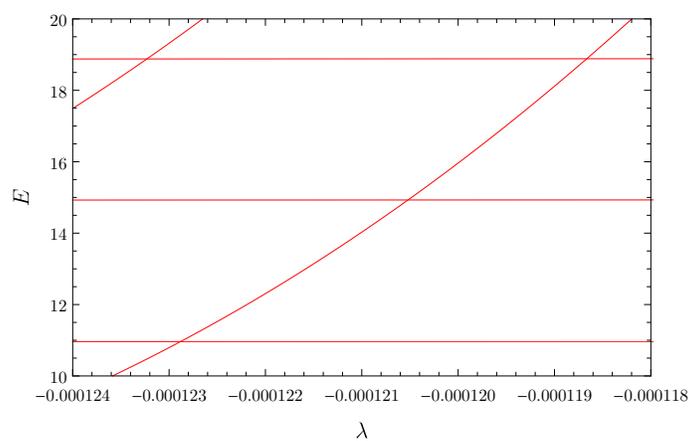}
\end{center}
\caption{Part of the spectrum of odd states of the anharmonic oscillator (%
\ref{eq:H_ANHO})}
\label{Fig:DETOD)}
\end{figure}

Figure~\ref{Fig:SPECTDET} shows that $E_{0}$ and $E_{1}$ remain isolated and
become eigenvalues of the harmonic oscillator when $\lambda \rightarrow
0^{-} $. The reason is that they are below the minima of the side
potentials. The eigenvalues $E_{2}$, $E_{3}$ and $E_{4}$ approach each other
and become quasi degenerate for intermediate values of $\lambda $. As $%
\lambda \rightarrow 0^{-}$ $E_{2}$ tends to a harmonic-oscillator eigenvalue
while the pair $\left( E_{3},E_{4}\right) $ remains quasi degenerate and
moves upwards. When $E_{3}$ meets $E_{5}$ there is an avoided crossing after
which $E_{3}$ approaches a harmonic oscillator eigenvalue while $E_{5}$
deviates upwards. The same situation takes place between $E_{4}$ and $E_{6}$%
, the former becomes a harmonic oscillator eigenvalue and the latter moves
upwards. All the higher eigenvalues follow the same pattern; for example, $%
E_{4k+1}$, $k=1,2,\ldots $, remain isolated till they are pushed upwards by
a lower odd-parity eigenvalue. The eigenvalues $(E_{4k+2},E_{4k+3},E_{4k+4})$%
, $k=0,1,\ldots $, become quasi degenerate at intermediate values of $%
\lambda $ before the pair $(E_{4k+3},E_{4k+4})$ separates and moves upwards.
It seems that every eigenvalue $E_{n}$ with $n>1$ undergoes an avoided
crossing with a higher eigenvalue of the same symmetry before becoming a
harmonic-oscillator eigenvalue. If $n>3$ the eigenvalue $E_{n}$ undergoes
avoided crossings with $E_{n-2}$ and $E_{n+2}$ as illustrated in the more
detailed figures \ref{Fig:DETEV} and \ref{Fig:DETOD)}. The eigenvalues
approach so closely that the avoided crossings appear actual crossings.

\begin{figure}[tbp]
\begin{center}
\bigskip\bigskip\bigskip \includegraphics[width=9cm]{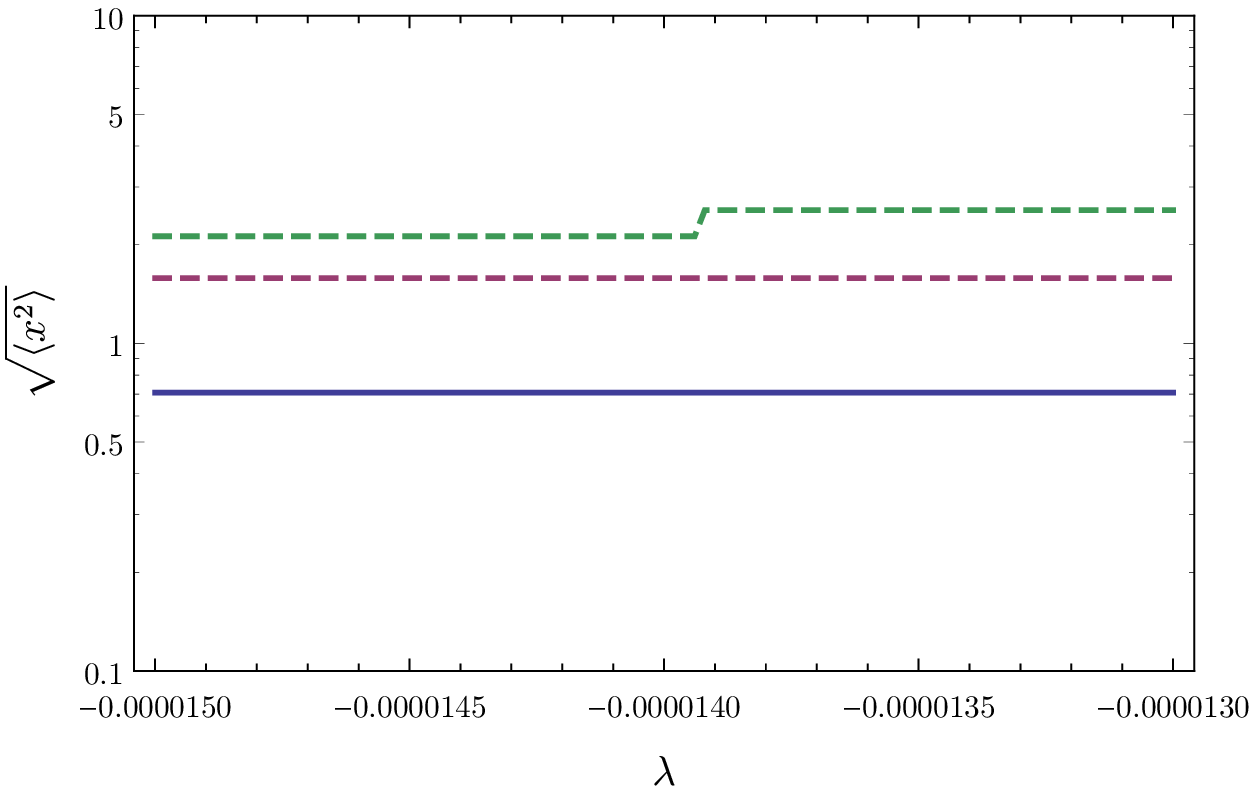}
\end{center}
\caption{$\protect\sqrt{\langle x^2\rangle}$ for the states of the
anharmonic oscillator (\ref{eq:H_ANHO}) with quantum numbers $n=0,2,4$}
\label{Fig:sq_x2_024)}
\end{figure}

\begin{figure}[tbp]
\begin{center}
\bigskip\bigskip\bigskip \includegraphics[width=9cm]{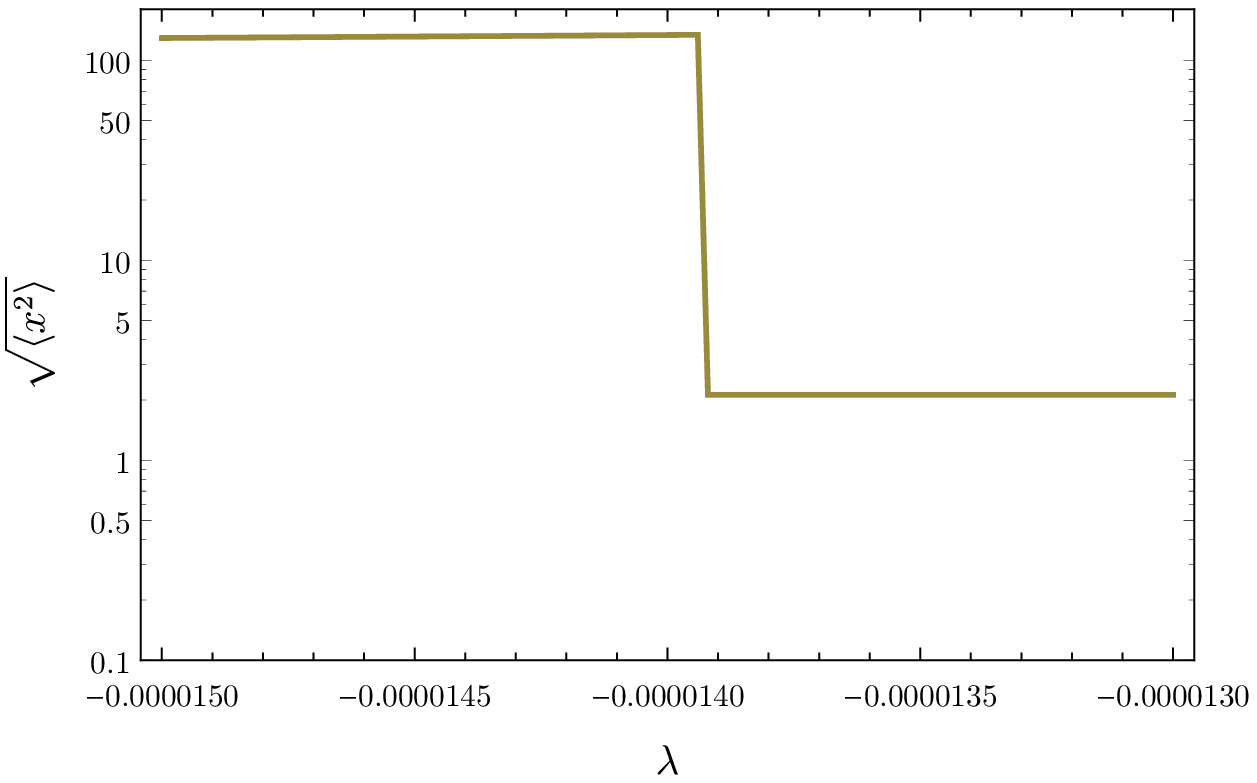}
\end{center}
\caption{$\protect\sqrt{\langle x^2\rangle}$ for the state of the anharmonic
oscillator (\ref{eq:H_ANHO}) with quantum number $n=6$}
\label{Fig:sq_x2_6)}
\end{figure}

In order to understand what happens at the avoided crossings we calculated $%
\Delta x=\sqrt{\left\langle x^{2}\right\rangle }$ for some states. This
root-mean-square deviation is expected to be larger when the state is
localized on the side wells. Figures \ref{Fig:sq_x2_024)} and \ref
{Fig:sq_x2_6)} show $\Delta x$ for the states with quantum numbers $%
n=0,2,4,6 $. The states $n=0,2$ do not participate in avoided crossings and
the corresponding $\Delta x$ does not change considerably as $\lambda
\rightarrow 0^{-}$. The state $n=4$ undergoes an avoided crossing and $%
\Delta x$ exhibits a jump that suggests that it changes from being localized
mainly on the central well to being localized mainly on the side ones. On
the other hand, the state $n=6$ appears to be mainly localized on the side
wells before the avoided crossing and mainly on the central one after it. In
this case the jump is considerably larger indicating that the form of the
eigenfunction changes more dramatically.

\section{Conclusions}

\label{sec:conclusions}

We revisited an old but interesting problem in quantum mechanics and
mathematical physics. It has been our purpose to outline some remarkable
features of its eigenvalues and eigenfunctions that have not been pointed
out before. In particular, the spectrum for $\lambda <0$ exhibits a rich
structure of avoided crossings at which the states that take part undergo
dramatic changes in their form. Such changes are clearly revealed by the
behaviour of the expectation value $\left\langle x^{2}\right\rangle (\lambda
)$. We also estimated the asymptotic behaviour of the coefficients of the
perturbation series and showed that they can be summed by means of Pad\'{e}
approximants and Borel-Pad\'{e} transformations for $\lambda >0$. This
calculation was greatly facilitated by the hypervirial perturbation method
that leads to straightforward recurrence relations for the perturbation
corrections to the eigenvalues and expectation values $\left\langle
x^{2n}\right\rangle $\cite{F01}. At present we do not know if there is any
suitable approximation for $\lambda <0$. In this region we simply resorted
to the Rayleigh-Ritz variational method with a basis set of $1000$
eigenfunctions of the harmonic oscillator. The reason is that the three
widely separated wells pose a quite difficult problem for accurate
calculation of the eigenfunctions and eigenvalues. We expect that present
investigation may be a suitable complement to previous ones about this
problem\cite{HS78,T16}

\begin{thebibliography}{9}
\bibitem{HS78}  I. W. Herbst and B. Simon, Phys. Lett. B 78 (1978) 304-306.
See also erratum Phys. Lett. B 80 (1979) 433.

\bibitem{T16}  A. V. Turbiner, Phys. Rep. 642 (2016) 1-71.

\bibitem{F01}  F. M. Fern\'{a}ndez, Introduction to Perturbation Theory in
Quantum Mechanics, (CRC Press, Boca Raton, 2001).

\bibitem{ZJ03}  J. Zinn-Justin, Ann. Inst. Fourier, Grenoble 54 (2003)
1259-1285.
\end{thebibliography}
\end{document}